\documentstyle[12pt,epsfig]{article}
\def\beq{\begin{equation}}
\def\eeq{\end{equation}}
\def\bea{\begin{eqnarray}}
\def\eea{\end{eqnarray}}
\def\nn{\nonumber}

\def\pl#1#2#3{{\it Phys. Lett. }{\bf B#1 }(19#2) #3}
\def\zp#1#2#3{{\it Z. Phys. }{\bf C#1 }(19#2) #3}

\def\pr#1#2#3{{\it Phys. Rev. }{\bf D#1 }(19#2) #3}
\def\np#1#2#3{{\it Nucl. Phys. }{\bf B#1 }(19#2) #3}

\def\sjnp#1#2#3{{\it Sov. J. Nucl. Phys. }{\bf #1 }(19#2) #3}

\begin{document}
\topmargin -1.0cm
\oddsidemargin -0.8cm
\evensidemargin -0.8cm
\pagestyle{empty} 
\begin{flushright}
CERN-TH/96-206\\
FTUV 96/48 - IFIC 96/56
\end{flushright}
\vspace*{6mm}
\begin{center}
{\Large\bf Next-to-leading order renormalization\\
 of the $\Delta B=2$ operators in the static theory}
\end{center}
\vspace{1cm}

\begin{center}
{\bf M.~Ciuchini}\footnote{On leave of absence from INFN,
Sezione Sanit\`a, V.le Regina Elena 299, 00161 Rome, Italy},
{\bf E.~Franco}\footnote{On leave of absence from INFN, Sezione di Roma I,
Dip. di Fisica, Univ. ``La Sapienza", Rome, Italy}\\
{\small Theory Division, CERN, 1211 Geneva 23, Switzerland}\\
\vspace{0.2cm}
and\\
\vspace{0.2cm}
{\bf V.~Gim\'enez}\\
{\small Dep. de Fisica Teorica and IFIC, Univ. de Valencia,}\\
{\small Dr. Moliner 50, E-46100, Burjassot, Valencia, Spain}\\
\end{center}
\vspace{2.5cm}
\centerline{\bf Abstract}

The renormalization, at the next-to-leading order
in $\alpha_s$, of the $\Delta B=2$ operators at the lowest order in the
heavy quark expansion, namely in the static theory, is computed taking
into account previously missed contributions. These operators are relevant
for the calculation of the $B^0$--$\bar B^0$ mixing on the lattice.

\vfill
\begin{flushleft}
CERN-TH/96-206\\
FTUV 96/48 - IFIC 96/56\\
July 1996
\end{flushleft}
\eject
~\vfill\eject
\setcounter{page}{1}
\setcounter{footnote}{0}
\baselineskip20pt
\pagestyle{plain}
 
\section{Introduction}

The lack of a precise knowledge of the matrix elements of the operator
\beq
\langle \bar B\vert \bar b\gamma^\mu(1-\gamma_5)d~\bar b\gamma_\mu(1-\gamma_5)d
\vert B\rangle=\frac{4}{3}f_B^2B_BM_B,
\label{eq:fb}
\eeq
generated by the box diagram with the exchange of virtual top quarks, is the
main source of theoretical error in the extraction of the
CKM matrix element $V_{td}$ from the $B^0$--$\bar B^0$ mixing parameter $x_d$.
This uncertainty on $f_B^2B_B$ propagates to other theoretical estimates.
In particular the extraction of the CP-violating phase from the combined
analysis of $x_d$ and the $K^0$--$\bar K^0$ CP-violating parameter
$\epsilon$ is
affected by a twofold ambiguity that would be eliminated by a precise
determination of $f_B^2B_B$. In turn this would strongly reduce the
uncertainty in the prediction of the asymmetry in the decay $B\to J/\psi~K_S$
\cite{cfmrs}.

The matrix element in eq.~(\ref{eq:fb}) can be evaluated on the lattice
\cite{latcal}. However, since $m_b$ is larger than the current values
of the lattice cutoff, the $b$ quark cannot be put on the lattice as a 
dynamical field. Therefore an effective theory based on the expansion in the
heavy quark mass is needed. Such a theory has been built \cite{hqt}, and its
discretized version can be used in lattice simulations. In particular the
expansion at the lowest order in $m_b$ is used to build the static effective
theory both in the continuum and on the lattice. This theory has the usual
form
\beq
{\cal H}_{static}=\sum_i C_i Q_i
\eeq
in terms of Wilson coefficients and local four-fermion operators.
In order to use lattice
results, the effective theory in the continuum must be matched both to its
lattice counterpart and to the ``full'' theory, namely a theory with a
dynamical heavy field. In ref.~\cite{fhh} this matching has been computed at
$O(\alpha_s)$, using the $\Delta B=2$ effective Hamiltonian \cite{bjw} as
the ``full" theory.

The two matching procedures take place at different scales. In fact the
matching to the ``full'' theory is done at a scale of the order of the
ultraviolet cutoff, namely $m_b$, while the matching to the static
theory on the lattice is done at typical current values of the lattice
cutoff, $1/a\sim 2$ GeV. A complete determination of the static theory in
the continuum requires the calculation of the running of the Wilson
coefficients between these two scales. This is usually done by using the
renormalization group equations (RGEs). In this way, at the leading order (LO),
one resums in the Wilson coefficients terms such as $\alpha_s^n \log^n$,
assumed to be of $O(1)$. To be
consistent, an $O(\alpha_s)$ matching calls for a
next-to-leading (NLO) determination of the Wilson coefficients, which resums
terms of the type $\alpha_s^{n+1} \log^n$.
Of course, in the case at hand, one may argue that the relevant logs, namely
$\log(a^2m_b^2)\sim 1.6$, are not large, so that the running at the leading
order can be considered as a pure $O(\alpha_s)$ effect. Anyway the calculation
of the anomalous dimension up to the NLO is required to have a
regularization-scheme-independent expression of the Wilson coefficients.
In the past an effort has been made to calculate this NLO anomalous
dimension in the static theory \cite{vg}. However, as
pointed out in ref.~\cite{bbl}, some contributions coming from the operator
mixing have been overlooked in ref.~\cite{vg}.

In the next section we calculate these new contributions and present a complete
determination of the NLO Wilson coefficients of the static theory.

\section{NLO Wilson coefficients in the static theory}

In this section we discuss the NLO gluon renormalization of the
$\Delta B=2$ operators at the lowest order in the heavy quark expansion,
i.e. in the static limit $m_b\to\infty$. We want to calculate the NLO
expression of the Wilson coefficients of the relevant operators. 
To this end a few steps are required. First of all, the basis of local
operators in the effective theory must be identified. Then the effective
theory has to be
matched against a ``full'' theory at a scale of the order of the ultraviolet
cutoff, fixing in this way the initial conditions of the renormalization
group equations. Finally the anomalous dimension matrix in the effective
theory must be calculated at the desired order in $\alpha_s$ and the RGE
solved to give the Wilson coefficients
as functions of the renormalization scale $\mu^2$. The scale-independent
physical amplitude is given by the product of the Wilson coefficients
$\vec C(\mu^2)$ and the matrix elements of the corresponding renormalized
operators $\langle \vec Q(\mu^2)\rangle$, the latter usually requiring some
non-perturbative method to be evaluated.

The $\Delta B=2$ operator basis in the static limit is given by two
dimension-six local operators
\bea
\vec Q&=&\left(
\begin{tabular}{c}
$Q_1$\\
$Q_2$
\end{tabular}\right)~,\nn\\
Q_1 = 2 \bar h^{(+)} \gamma_\mu (1-\gamma_5)d~
\bar h^{(-)} \gamma^\mu (1-\gamma_5) d~,&~&
Q_2 = 2 \bar h^{(+)} (1-\gamma_5) d~ \bar h^{(-)} (1-\gamma_5) d~,
\label{eq:basis}
\eea
where the field $\bar h^{(+)}$ creates a heavy quark and $\bar h^{(-)}$
annihilates a heavy antiquark. Our calculation explicitly shows that indeed
the basis is closed under renormalization at the NLO.

The next step is the NLO matching. The heavy quark theory has to be matched
against the $\Delta B=2$ effective Hamiltonian by comparing
the matrix elements at the scale $m_b$ of the relevant operators in the
``full'' and the effective theories, up to and including $O(\alpha_s)$ terms.
Here the effective Hamiltonian plays the role of the ``full'' theory, even if
it is also an effective theory with the top and the heavy bosons integrated out,
which in turn is matched against the Standard Model at the weak scale. The 
$\Delta B=2$ effective Hamiltonian has been calculated in ref. 
\cite{bjw} and is completely known at the NLO. There exists only one
$\Delta B=2$ operator in this theory, namely
\beq
Q_{LL}=\bar b \gamma_\mu (1-\gamma_5) d~\bar b \gamma^\mu (1-\gamma_5) d~.
\eeq
The calculation of the matching of this operator onto the operators in
eq.~(\ref{eq:basis}) requires the expansion of the matrix element
$\langle b\bar d\vert Q_{LL}\vert \bar b d\rangle$ in the heavy quark mass.
This calculation has been done in refs. \cite{fhh,vg} and gives the initial
condition at the scale $m_b$
\bea
&~&\vec C(m_b^2)=\left(
\begin{tabular}{c}
$1+\frac{\alpha_s(m_b^2)}{4\pi}B_1$\\
$\frac{\alpha_s(m_b^2)}{4\pi}B_2$
\end{tabular}\right)C_{LL}(m_b^2),\nn\\
&~&B_1=\left\{
\begin{tabular}{ll}
$-14$ & NDR\\
$-11$ & DRED
\end{tabular}\right.,~~B_2=-8~.
\eea
We have reported the values of $B_1$ both in the naive dimensional
regularization
scheme (NDR) and in dimensional reduction (DRED). This scheme dependence
cancels out against the corresponding dependence contained in the Wilson
coefficient $C_{LL}(m_b^2)$ of
the $\Delta B=2$ effective Hamiltonian\footnote{As noticed in ref.~\cite{vg},
this implies that the anomalous dimension matrix
element $\hat\gamma^{(1)}_{11}$, see eq.~(\ref{eq:exp}), in the effective
theory is the same in DRED and NDR.}.
The other initial condition, $C_2(m_b^2)$, is the same both in DRED and
NDR. We notice that, starting at $O(\alpha_s)$, it does not pick up
the scheme-dependent part inside $C_{LL}(m_b^2)$ at the NLO.

\begin{figure}[t]
\centering
\epsfysize=0.45\textheight
\leavevmode\epsffile{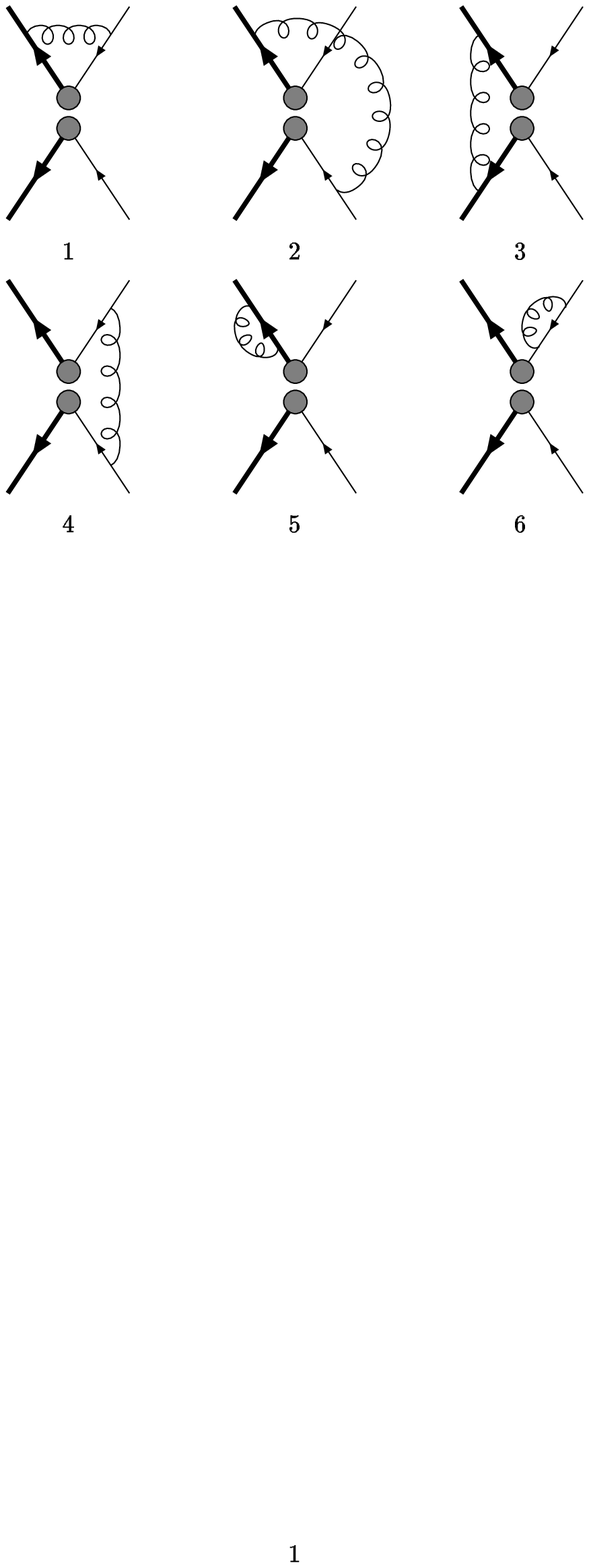}
\caption[]{\it{The Feynman diagrams that contribute to $\hat\gamma^{(0)}$.
Thick (thin) lines represent the heavy (light) quarks. The blobs are the
operator insertion point.}}
\label{fig:diags}
\end{figure}

The evolution of the Wilson coefficients between the
matching scale $m_b^2$ and the renormalization scale $\mu^2$ is determined by
the renormalization group equation
\beq
\mu^2\frac{d}{d\mu^2}\vec C(\mu^2)=\frac{1}{2}\hat\gamma^T \vec C(\mu^2)~.
\label{eq:rge}
\eeq
The anomalous dimension matrix $\hat\gamma$ is defined as
\beq
\hat\gamma=2\mu^2\frac{d}{d\mu^2}\hat Z~,
\eeq
where $\hat Z$ is the matrix of the renormalization constants connecting
the bare and the renormalized operators
\beq
\vec O_R=\hat Z^{-1} \vec O_B~.
\eeq
The formal solution of eq.~(\ref{eq:rge}) is
\beq
\vec C(\mu^2)=T_{\alpha_s}e^{\int_{\alpha_s(m_b^2)}^{\alpha_s(\mu^2)}
d\alpha_s \frac{\hat\gamma(\alpha_s)}{2\beta(\alpha_s)}}\vec C(m_b^2)~,
\eeq
where $T_{\alpha_s}$ is the ordered product with increasing powers of
the coupling constant from left to right and $\beta(\alpha_s)$ is the
QCD beta function
\beq
\beta(\alpha_s)=\mu^2\frac{d\alpha_s}{d\mu^2}~.
\eeq

In order to calculate the Wilson coefficients at the NLO, the first
two terms of the perturbative expansion of $\beta(\alpha_s)$ and
$\hat\gamma(\alpha_s)$ are needed:
\bea
\beta(\alpha_s)&=&-\frac{\alpha_s^2}{4\pi}\beta_0-\frac{\alpha_s^3}{(4\pi)^2}
\beta_1+\dots~,\nn\\
\hat\gamma(\alpha_s)&=&\frac{\alpha_s}{4\pi}\hat\gamma^{(0)}+\left(
\frac{\alpha_s}{4\pi}\right)^2\hat\gamma^{(1)}+\dots~.
\label{eq:exp}
\eea
The beta function coefficients are well known:
\beq
\beta_0=\frac{11N-2n_f}{3}~,\qquad\beta_1=\frac{34}{3}N^2-\frac{10}{3}Nn_f
-2 C_F n_f~,
\label{eq:beta}
\eeq
where $N$ is the number of colours, $n_f$ is the number of active flavours and
$C_F=(N^2-1)/2N$. The anomalous dimension matrices $\hat\gamma^{(0)}$ and
$\hat\gamma^{(1)}$ are calculated by computing the one- and two-loop
renormalization of the operators in eq.~(\ref{eq:basis}).

\begin{table}[t]
\begin{center}
\begin{tabular}{|c|c|c|c|c|c|} 
\hline
Diag. & Mult. & $Q_1\to Q_1$ & $Q_1\to Q_2$ & $Q_2\to Q_1$ & $Q_2\to Q_2$ \\
\hline
1 & 2 & $2\xi C_F$ & - & - & $2\xi C_F$ \\
2 & 2 & $\xi\left(1-\frac{1}{N}\right)$ & - & $-\frac{\xi}{2}$ &
$-\xi\left(1+\frac{1}{N}\right)$ \\
3 & 1 & $\frac{3-\xi}{2}\left(1-\frac{1}{N}\right)$ & - &
$-\frac{3-\xi}{4}$ & $-\frac{3-\xi}{2}\left(1+\frac{1}{N}\right)$ \\
4 & 1 & $-\frac{3+\xi}{2}\left(1-\frac{1}{N}\right)$ & - &
$\frac{1+\xi}{4}-\frac{1}{2N}$ & $-\frac{1-\xi}{2}
\left(1+\frac{1}{N}\right)$ \\
5 & 1 & $\left(3-\xi\right)C_F$ & - & - & $\left(3-\xi\right)C_F$ \\
6 & 1 & $-\xi C_F$ & - & - & $-\xi C_F$ \\
\hline
\end{tabular}
\caption[]{\it{Pole coefficients of the operator insertions into the
diagrams of fig.~\ref{fig:diags}, calculated in the linear $R_\xi$ gauge.
An overall factor $\alpha_s/4 \pi$ is understood.
The second column contains the diagram
multiplicity factors, which have already been applied to the shown
coefficients. To account for the renormalization of the external legs,
self-energy diagrams count as $1/2$.}}
\label{tab:poles}
\end{center}
\end{table}

The one-loop renormalization matrix is given by the infinite parts of the
operator insertion into the diagrams in fig.~\ref{fig:diags}, computed in
the static theory. We have used the dimensional regularization\footnote{
$\gamma_5$ and subtraction prescriptions are immaterial for the anomalous
dimension at the LO.} to calculate the divergent parts of these
diagrams that appear as poles in $\epsilon=(4-D)/2$. The coefficients of
the poles are collected in table~\ref{tab:poles}. The calculation is
straightforward, the only peculiarity being that few tensor structures
can appear in the effective theory because of the equation satisfied by
the static fields, $v_\mu \gamma^\mu h^{(\pm)}=\pm h^{(\pm)}$. In
particular tensors can be reduced in the following way:
\bea
\bar h^{(+)} \sigma_{\mu\nu}(1-\gamma_5) d~\bar h^{(-)} \sigma^{\mu\nu}
(1-\gamma_5) d &=& 4\Bigl[\bar h^{(+)} \gamma^\mu(1-\gamma_5) d~\bar h^{(-)}
\gamma_\mu(1-\gamma_5) d\nn\\
&~&+\bar h^{(+)}(1-\gamma_5) d~\bar h^{(-)}(1-\gamma_5)d\Bigr]~.
\eea

The anomalous dimension at the LO is minus twice the pole coefficients in the
matrix $\hat Z$ of the operator renormalization constants, so that, from
table~\ref{tab:poles}, we readily find
\beq
\hat\gamma^{(0)}=\left(
\begin{tabular}{c c}
$-6 C_F$ & 0\\
$1+\frac{1}{N}$ & $-6C_F+4+\frac{4}{N}$
\end{tabular}\right)~.
\label{eq:gamma0}
\eeq
As expected, the anomalous dimension matrix is independent of the gauge.
$\hat\gamma^{(0)}_{11}$ agrees with the previous
calculations \cite{vs}, while the other matrix elements were not
considered previously.
The form of this matrix has some important consequences. On the one
hand, $Q_1$ insertion has vanishing component on $Q_2$ at the leading order.
Since the initial condition of $Q_2$ is already of order $\alpha_s$, the
one-loop contribution of $Q_2$ becomes a NLO effect. For the very
same reason, we do not need to calculate the two-loop renormalization,
i.e. the second row of $\hat\gamma^{(1)}$, which would generate a
next-to-next-to-leading order term. On the other hand, $Q_2$ has 
a non-zero leading component on $Q_1$, and thus contributes to the Wilson
coefficient $C_1$ at the NLO.

We still have to calculate the first row of $\hat\gamma^{(1)}$. This is a
hard task, involving the evaluation of the pole parts of several two-loop
diagrams. However the renormalization of $Q_1$ onto itself has already
been calculated in ref.~\cite{vg}, while the insertion of $Q_1$
has vanishing component onto $Q_2$. This last statement holds to all orders
in perturbation theory, provided that one chooses a renormalization scheme that
preserves the Fierz symmetry. In fact
\beq
Q^{(+)}=Q_1~,\qquad Q^{(-)}=Q_2+\frac{1}{4}Q_1
\eeq
are the eigenvectors of the Fierz transformation with eigenvalues $\pm 1$
\footnote{This is a consequence of the properties of the static fields,
$v_\mu\gamma^\mu h^{(\pm)}=\pm h^{(\pm)}$.}, therefore they cannot mix under
renormalization. This enforces the following relations among the anomalous
dimension matrix elements
\beq
\hat\gamma_{12}=0~,\qquad \hat\gamma_{21}=\frac{1}{4}\Bigl(\hat\gamma_{22}
-\hat\gamma_{11}\Bigr)~,
\eeq
which indeed are satisfied by eq.~(\ref{eq:gamma0}).

The NLO anomalous dimension matrix is then given by
\bea
\label{eq:gamma1}
&~&\hat\gamma^{(1)}=\\
&~&\left(
\begin{tabular}{c c}
$-\frac{N-1}{12N}\left[127N^2+143N+63-\frac{57}{N}+8\pi^2
\left(N^2-2N+\frac{4}{N}\right)-n_f\left(28N+44\right)\right]$ & 0\\
$\frac{1}{4}\Bigl(X-\hat\gamma^{(1)}_{11}\Bigr)$ & $X$\nn

\end{tabular}\right)~,
\eea
where, as already noticed, the entry marked with $X$ is not needed at
the NLO.

We are now ready to write down the solution of the RGE, eq.~(\ref{eq:rge}),
at the NLO. Using eqs.~(\ref{eq:beta}), (\ref{eq:gamma0}) and
(\ref{eq:gamma1}), we obtain
\bea
C_1(\mu^2)&=&\left(\frac{\alpha_s(m_b^2)}{\alpha_s(\mu^2)}\right)^{d_1}
\left(1+\frac{\alpha_s(\mu^2)-\alpha_s(m_b^2)}{4\pi} J\right)
C_1(m_b^2)\nn\\
&~&+\left[\left(\frac{\alpha_s(m_b^2)}{\alpha_s(\mu^2)}\right)^{d_2}-
\left(\frac{\alpha_s(m_b^2)}{\alpha_s(\mu^2)}\right)^{d_1}\right]
\frac{\hat\gamma^{(0)}_{21}}{\hat\gamma^{(0)}_{22}-\hat\gamma^{(0)}_{11}}
C_2(m_b^2)\nn\\
C_2(\mu^2)&=&\left(\frac{\alpha_s(m_b^2)}{\alpha_s(\mu^2)}\right)^{d_2}
C_2(m_b^2)~,
\label{eq:wc}
\eea
where
\beq
d_i=\frac{\hat\gamma^{(0)}_{ii}}{2\beta_0}~,\qquad J=\beta_1\frac{d_1}
{\beta_0}-\frac{\hat\gamma^{(1)}_{11}}{2\beta_0}~.
\eeq
The new contribution to the NLO running of $C_1$ is the term proportional
to $\hat\gamma^{(0)}_{21}$, while $C_2$ is renormalized multiplicatively.

Numerically the new term contributes to the running between $m_b^2$ and
a typical lattice scale $\mu^2=4~\mbox{GeV}^2$ by increasing $C_1$ of a
few per cent, roughly doubling the already known NLO enhancement coming from
$\hat\gamma^{(1)}_{11}$. Moreover the operator $Q_2$ also contributes at
the NLO and should be included in lattice calculations of $f_B^2B_B$ in
the static limit.

\section*{Acknowledgements}
We thank G.~Martinelli and M.~Neubert for useful discussions
on the subject of this work. V.G.~wishes to acknowledge partial support by
CICYT under grant number AEN-96/1718.


\begin{thebibliography}{99}
\bibitem{cfmrs} M.~Lusignoli, L.~Maiani, G.~Martinelli and L.~Reina,
\np{369}{92}{139};\\
 M.~Ciuchini, E.~Franco, G.~Martinelli, L.~Reina and L.~Silvestrini,
\zp{68}{95}{239}.\bibitem{latcal} E.~Eichten, \np{~(Proc.~Suppl.)~4}{88}{170};\\
A.~Abada et al., \np{376}{92}{172};\\
A.~Soni, \np{~(Proc.~Suppl.)~47}{96}{43};\\
JLQCD Collaboration, S.~Aoki et al., \np{~(Proc.~Suppl.)~47}{96}{433};\\
UKQCD Co\-lla\-bo\-ra\-tion, A.K.~E\-wing et al.,
E\-DIN\-BURGH-95-550. To appear in {\it Phys. Rev.} {\bf D}
[hep-lat/9508030];\\
J.~Christensen, T.~Draper and C.~McNeile, poster presented 
at Lattice '96, St. Louis, USA. To be published in 
{\it Nucl.~Phys.}{\bf ~B (Proc.~Suppl.)};\\
T.~Draper and C.~McNeile,
\np{~(Proc.~Suppl.)~47}{96}{429};\\
V.~Gim\'enez et al., talk presented at Lattice '96, St. Louis, 
USA. To be published in {\it Nucl.~Phys.}{\bf ~B (Proc.~Suppl.)};\\
V.~Gim\'enez and G.~Martinelli, in preparation.
\bibitem{hqt} E.~Eichten and B.~Hill, \pl{234}{90}{511};\\
 H.~Georgi, \pl{240}{90}{447};\\
 B.~Grinstein, \np{339}{90}{253}.
\bibitem{fhh} J.M.~Flynn, O.F.~Hern\'andez and B.R.~Hill, \pr{43}{91}{3709}.
\bibitem{bjw} A.J.~Buras, M.~Jamin and P.H.~Weisz, \np{347}{90}{491}.
\bibitem{vg} V.~Gim\'enez, \np{401}{93}{116}.
\bibitem{bbl} G.~Buchalla, A.J.~Buras and M.E.~Lautenbacher, prep.
 MPI-Ph/95-104 [hep-ph/9512380].
\bibitem{vs} M.B.~Voloshin and M.A.~Shifman, \sjnp{45}{87}{292};\\
 H.D.~Politzer and M.B.~Wise, \pl{206}{88}{681}.
\end{thebibliography}
\end{document}